# A spectroscopic survey of the small near-Earth asteroid population: peculiar taxonomic distribution and phase reddening

*D. Perna*[(1,2)][1], *M. A. Barucci*[(2)], *M. Fulchignoni*[(2)], *M. Popescu*[(2,3)], *I. Belskaya*[(2,4)], *S. Fornasier*[(2)], *A. Doressoundiram*[(2)], *C. Lantz*[(2,5)], *F. Merlin*[(2)]

[(1)] INAF – Osservatorio Astronomico di Roma, Via Frascati 33, 00078 Monte Porzio Catone, Italy

[(2)] LESIA – Observatoire de Paris, PSL Research University, CNRS, Sorbonne Universités, UPMC Univ. Paris 06, Univ. Paris Diderot, Sorbonne Paris Cité, 5 place Jules Janssen, 92195 Meudon, France

[(3)] Astronomical Institute of the Romanian Academy, 5 Cuţitul de Argint, 040557 Bucharest, Romania

[(4)] Institute of Astronomy, Kharkiv V.N. Karazin National University, Sumska Str. 35, Kharkiv 61022, Ukraine

[(5)] Institut d'Astrophysique Spatiale, CNRS, UMR-8617, Université Paris-Sud, bâtiment 121, 91405 Orsay, France

## Abstract

We present the results of the first-ever visible spectroscopic survey fully dedicated to the small (absolute magnitude H≥20) near-Earth asteroid (NEA) population. Observations have been performed at the New Technology Telescope (NTT) of the European Southern Observatory (ESO), during a 30-night Guaranteed Time Observations programme, in the framework of the European Commission financed NEOShield-2 project. The visible spectra of 147 objects have been obtained and taxonomically classified. They show a peculiar taxonomic distribution, with respect to larger NEAs. In particular, olivine-rich A-types and organic-rich D-types are more abundant than what could be expected by extrapolating the taxonomic distribution of larger NEAs. Such results have implications for the investigation of the first phases of solar system history, including the delivery of prebiotic material on the early Earth. Having been obtained over a large range of solar phase angles, our data allowed us to evidence peculiar phase reddening behaviours for asteroids belonging to different taxonomic types. Low-albedo asteroids display no or limited phase reddening, compared to moderate- and high-albedo objects. This result suggests a promising novel way to distinguish primitive asteroids in the X-complex. In agreement with previous laboratory experiments, olivine-rich surfaces are the most affected by phase reddening.

[1] E-mail: davide.perna@oa-roma.inaf.it; davide.perna@obspm.fr

# 1 Introduction

The investigation of near-Earth asteroids (NEAs) can provide crucial information on the formation and early evolution of the solar system, including topics like the delivery of water and organic-rich material to the early Earth, and the emergence of life (e.g., Marty et al., 2016). The reserve in water and rare metals held by NEAs is also getting the attention from government agencies and private companies, as asteroid mining could expand the Earth's resource bases in the near future (e.g., Sanchez & McInnes, 2013). Furthermore, NEAs represent a well-founded threat to human beings and life in general: the PHAs ("potentially hazardous asteroids") could in principle collide with the Earth within the next century and cause extensive damage (e.g., Perna et al., 2013, 2016).

The discovery rate of NEAs is constantly increasing and rapidly approaching the 2000 objects/year barrier. Current discoveries mainly concern "small" (tens/hundreds of metres in diameter) NEAs close approaching the Earth. The investigation of such small-sized NEAs is particularly important to constrain the asteroidal contribution to the delivery of prebiotic material (water and organic molecules) to our planet (e.g., Pierazzo & Chyba, 1999; Morbidelli et al., 2000; Saladino et al., 2013; O'Brien et al., 2014). Also in terms of the current impact risk, the small objects deserve our particular attention, as they have the highest statistical likelihood of impact, and can still produce a catastrophe at a regional/national scale (e.g., Perna et al., 2015a). More in general, the proximity of NEAs allows us to study asteroids about two to three orders of magnitude smaller than those observable in the main belt (i.e., down to metre-sized objects), hence to open new frontiers in asteroid science. Indeed, recent results already evidenced that small asteroids behave differently than the larger bodies in terms of rotational properties (Statler et al., 2013) and regolith generation (Delbo et al., 2014).

However, such small asteroids become bright enough to be physically characterized from Earth only for very limited time spans, coinciding with their close approaches with our planet, whereupon they could be unobservable for years or even for decades. Rapid-response physical observations of such bodies are hence necessary in order to not leave the characterization rate behind the discovery rate. In particular, visible and near-infrared photometry and reflectance spectroscopy of asteroids allow their taxonomic classification and provide clues about their surface composition, mineralogy and scattering properties. Such techniques can be "calibrated" via laboratory measurements on minerals and meteorites, as well as on the first returned samples from asteroids that we are getting in these years (e.g., DeMeo et al., 2015; Reddy et al., 2015). Currently, the fraction of NEAs with assigned taxonomic class (with respect to the known population in the same size range) drops from ~1/3 for km-sized bodies to ~1/10 for objects in the 0.3-1 km range, to ~1/100 for NEAs smaller than 300 m.

To soften this deficiency, in the framework of the NEOShield-2 project[2], we performed the first-ever spectroscopic survey fully dedicated to the "small" NEAs, through a 30-night Guaranteed Time Observations programme at the 3.6-m New Technology Telescope of the European Southern Observatory (La Silla, Chile). The observational circumstances, as well as the data reduction and analysis are described in Section 2. In Section 3 we present our results, while Section 4 contains a discussion and our conclusions.

[2] http://www.neoshield.eu/

## 2 Methods

For our astronomical observations, the EFOSC2 spectrograph was used with the Grism #1 diffraction element. This configuration covers the spectral interval 0.40-0.92 μm with a resolution R~500 (~1.4 nm/pixel). Data were collected during twenty-four different nights spanning two years (April 2015 to March 2017), while six out of the thirty allocated nights were lost due to bad weather conditions. During each observing run, we gave preference to the smallest NEA observable (many of our targets were newly discovered bodies in proximity of their close approaches with the Earth), and limited our observations to objects with absolute magnitude equal or fainter than H=20 (corresponding to a maximum diameter of about 300 m assuming a value of 0.20 for the visual albedo). On each observing night, we also observed several solar analogue stars. All of the spectra have been acquired through the 2 arcsec slit, oriented along the parallactic angle to minimize the effects of atmospheric differential refraction.

The data reduction followed standard procedures (e.g., Perna et al., 2015b): bias and background subtraction, flat-field correction, one-dimensional spectra extraction, atmospheric extinction correction. We used the Octave and IRAF software packages. Wavelength calibration was obtained using He-Ar lamps emission lines. The reflectance of the asteroids (normalized at 0.55 μm) was obtained by dividing their spectra by those of solar analogue stars observed close in time and in airmass to the scientific frames.

We taxonomically classified each object by performing curve matching with the visible part of the 25 template spectra defined by the Bus-DeMeo scheme (DeMeo et al., 2009), using the M4AST online tool (Popescu et al., 2012).

Overall, in this work we report the observations of 147 "small" NEAs. For all of them, Table 1 lists the asteroid number/designation, the absolute magnitude *H*, the observational circumstances (date and UT starting time, exposure time, airmass, and phase angle), as well as the solar analogue star used to obtain the reflectance spectrum (and the airmass at which the star was observed). The table also includes the spectral slopes and the associated errors computed using the Octave's *polyfit* routine (no further statistical/systematic potential sources of errors are considered in our analysis) in the 0.44-0.65 μm range (this wavelength interval was selected to roughly correspond to the B-R photometric colour, and to avoid both the silicate band and the lower signal-to-noise region longward of ~0.7 μm), as well as the result of the taxonomic classification, the considered albedo (cf. Table notes) and the computed equivalent diameter of each object. For NEA 2011 AM24 we found a quite odd spectral shape, not fitting any of the taxonomic types: while we tentatively classify this object as a potential D-type, we discard it from our following analysis. The distribution of the equivalent diameters of our targets is shown in Fig. 1. The spectra of 129 asteroids are presented in Fig. 2, while the remaining spectra of 18 objects classified as belonging to the D or A taxonomic types are presented and discussed in Perna et al. (2017; regarding D-type asteroid 1993 HA), Barucci et al. (2018; regarding further 9 D-types) and Popescu et al. (2018; regarding A-types).

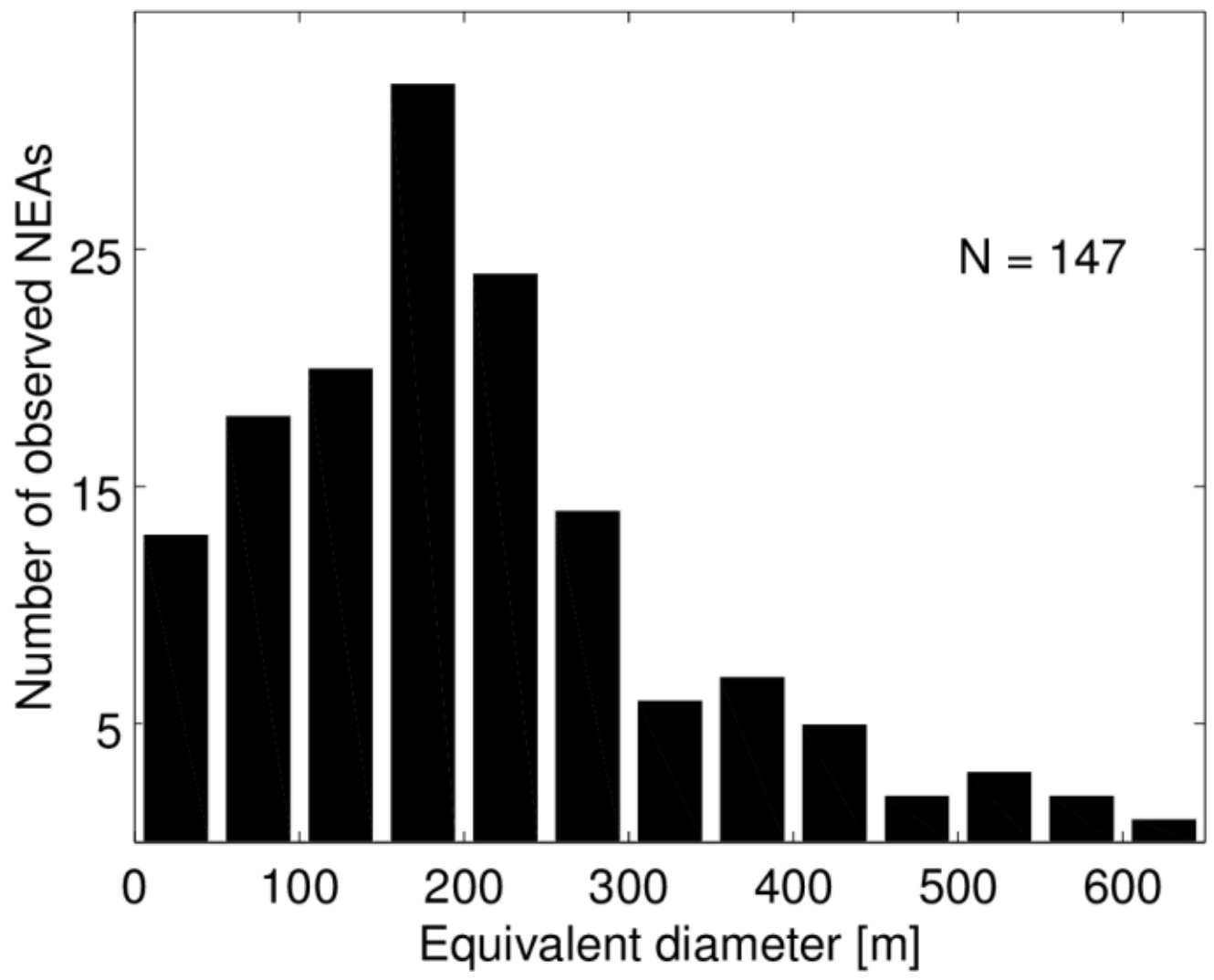

**Figure 1: Distribution of the equivalent diameters of the observed NEAs.**

**Table 1: Observational circumstances and data analysis results.**

| Object | H (mag) | Date/UT$_{start}$ | $t_{exp}$ (s) | Airmass | Solar analogue (airmass) | Phase angle (deg) | Slope (%/$10^3$ Å) | Taxon | Albedo | Equiv. diam. (m) |
|---|---|---|---|---|---|---|---|---|---|---|
| 4581 Asclepius | 20.7 | 2015-6-8 6:02 | 3×1200 | 1.097 | SA102-1081 (1.149) | 18.8 | 6.163±0.740 | Cgh | 0.065 | 378 |
| 52381 1993 HA | 20.1 | 2015-11-5 6:41 | 2×450 | 1.06 | HD11123 (1.021) | 41.1 | 12.365±0.338 | D | 0.140(1) | 339 |
| 138852 2000 WN10 | 20.2 | 2015-11-6 8:15 | 2×450 | 1.091 | HD9729 (1.089) | 57.3 | 11.649±0.267 | Sq | 0.243 | 246 |
| 162783 2000 YJ11 | 20.6 | 2015-7-19 2:19 | 1×1200 | 1.109 | SA112-1333 (1.366) | 17.7 | 16.642±1.019 | S | 0.211 | 219 |
| 163348 2002 NN4 | 20 | 2016-8-29 0:49 | 2×1200 | 1.187 | HD202282 (1.029) | 44.5 | 3.151±0.476 | X | 0.047 | 613 |
| 164202 2004 EW | 20.7 | 2015-6-9 9:20 | 2×600 | 1.082 | Hip106102 (1.051) | 44 | 5.167±0.526 | X | 0.340(2) | 165 |
| 216523 2001 HY7 | 20.5 | 2015-4-14 9:41 | 2×600 | 1.176 | SA112-1333 (1.260) | 34.6 | 8.480±1.057 | Sq | 0.243 | 214 |
| 281375 2008 JV19 | 20.7 | 2015-7-20 5:40 | 1×900 | 1.369 | SA110-361 (1.176) | 39.4 | 3.356±0.578 | C | 0.050 | 431 |
| 293726 2007 RQ17 | 22.5 | 2015-6-8 7:44 | 2×600 | 1.288 | SA107-998 (1.265) | 50.7 | 40.851±0.477 | A | 0.191 | 96 |
| 307070 2002 AV31 | 20.6 | 2017-1-3 7:54 | 2×1000 | 1.222 | SA102-1081 (1.143) | 33.8 | 5.899±0.398 | Xk | 0.095 | 327 |
| 307564 2003 FQ6 | 20.8 | 2016-8-30 4:45 | 3×1200 | 1.084 | HD202282 (1.038) | 27.1 | 13.903±0.236 | S | 0.211 | 200 |
| 326302 1998 VN | 20.5 | 2016-11-29 8:07 | 1×800 | 1.145 | HD33792 (1.191) | 94.1 | 14.560±0.688 | V | 0.362 | 175 |
| 330659 2008 GG2 | 22.8 | 2016-4-1 6:07 | 1×1100 | 1.05 | HD171207 (1.030) | 12.2 | 13.845±1.567 | Sq | 0.243 | 74 |
| 334412 2002 EZ2 | 20.2 | 2016-4-1 4:48 | 2×700 | 1.224 | HD106649 (1.039) | 8.1 | 13.979±0.336 | S | 0.400(2) | 192 |
| 350713 2001 XP88 | 20.7 | 2015-7-19 8:10 | 2×900 | 1.191 | SA115-271 (1.290) | 15.4 | 7.062±0.303 | Xc | 0.129 | 268 |
| 388945 2008 TZ3 | 20.4 | 2016-4-1 6:25 | 2×600 | 1.343 | SA107-998 (1.358) | 25.6 | 4.250±0.285 | C | 0.050 | 494 |
| 401885 2001 RV17 | 20.4 | 2015-11-6 1:40 | 2×1100 | 1.244 | SA112-1333 (1.260) | 84.2 | 15.737±0.480 | Sv | 0.309 | 199 |
| 410777 2009 FD | 22.1 | 2015-11-7 0:39 | 2×900 | 1.53 | HD9729 (1.452) | 39.7 | 4.326±0.257 | X | 0.010(3) | 505 |
| 418849 2008 WM64 | 20.7 | 2015-12-14 7:11 | 1×900 | 1.415 | SA98-978 (1.371) | 73.8 | 18.279±0.376 | Sa | 0.367 | 159 |
| 420262 2011 KD11 | 20.1 | 2015-6-9 0:31 | 2×1050 | 1.024 | SA102-1081 (1.556) | 30.4 | 18.345±0.636 | D | 0.048 | 579 |
| 420738 2012 TS | 20.8 | 2015-11-6 5:42 | 2×900 | 1.037 | HD36649 (1.125) | 51.6 | 7.763±0.345 | X | 0.047 | 424 |
| 433953 1997 XR2 | 20.8 | 2017-1-2 3:00 | 2×630 | 1.209 | SA102-1081 (1.142) | 31 | 8.701±0.314 | Xe | 0.136 | 249 |
| 438955 2010 LN14 | 21.1 | 2015-6-8 9:02 | 2×600 | 1.065 | SA102-1081 (1.149) | 35.5 | 7.418±0.262 | Q | 0.227 | 168 |
| 441987 2010 NY65 | 21.5 | 2015-7-19 1:14 | 2×1200 | 1.226 | SA107-998 (1.269) | 46 | 19.824±0.491 | Sv | 0.071(4) | 250 |
| 444584 2006 UK | 20.2 | 2016-5-12 6:28 | 1×600 | 1.731 | SA107-998 (1.313) | 29.2 | 25.838±0.961 | A | 0.191 | 277 |
| 446924 2002 VV17 | 20.1 | 2015-11-6 6:12 | 2×300 | 1.454 | SA98-978 (1.144) | 14.1 | 9.537±0.256 | Q | 0.227 | 266 |
| 447221 2005 UO5 | 20.7 | 2015-11-6 3:54 | 2×750 | 1.087 | HD9729 (1.089) | 43.8 | 8.022±0.519 | Q | 0.227 | 202 |
| 452302 1995 YR1 | 20.3 | 2015-12-14 6:49 | 1×600 | 1.347 | SA98-978 (1.371) | 35.3 | 7.315±0.256 | Xk | 0.095 | 376 |
| 455659 2005 BO1 | 21.6 | 2015-12-15 4:40 | 1×900 | 1.111 | SA98-978 (1.145) | 28 | 9.782±0.848 | Q | 0.227 | 134 |
| 457663 2009 DN1 | 20.3 | 2016-7-1 8:09 | 2×1200 | 1.196 | SA102-1081 (1.278) | 55.5 | 3.041±0.598 | X | 0.047 | 534 |
| 458135 2010 GE25 | 20.2 | 2016-3-30 1:54 | 1×1200 | 1.256 | SA102-1081 (1.297) | 62.4 | 11.218±0.928 | Sv | 0.230(4) | 253 |
| 462559 2009 DD1 | 20.4 | 2016-8-30 1:32 | 2×1200 | 1.231 | SA110-361 (1.229) | 46.4 | 10.121±0.655 | Xe | 0.136 | 300 |
| 466507 2014 FK33 | 20.8 | 2016-4-1 7:13 | 2×850 | 1.135 | HD171207 (1.030) | 13.2 | 12.714±0.808 | Sq | 0.243 | 187 |
| 467352 2003 KZ18 | 21.3 | 2015-6-9 2:40 | 2×900 | 1.065 | Hip56139 (1.030) | 32.4 | 9.050±0.465 | D | 0.048 | 333 |
| 468005 2012 XD112 | 21.2 | 2015-12-14 2:24 | 1×1200 | 1.325 | SA98-978 (1.371) | 24.8 | 9.373±0.702 | Sq | 0.243 | 155 |
| 468681 2009 MZ6 | 20.5 | 2016-7-1 5:23 | 1×1200 | 1.228 | SA102-1081 (1.278) | 19.8 | 7.768±1.005 | Q | 0.227 | 222 |
| 468741 2010 VM1 | 20.1 | 2016-7-1 7:13 | 2×1200 | 1.326 | HD206938 (1.240) | 42.3 | 11.777±0.364 | Q | 0.227 | 266 |
| 469737 2005 NW44 | 20.4 | 2016-6-30 8:31 | 1×900 | 1.094 | HD146997 (1.240) | 49.3 | 8.502±0.724 | Xe | 0.136 | 300 |
| 470864 2008 YV148 | 20.5 | 2016-6-30 7:18 | 2×1800 | 1.022 | HD220764 (1.045) | 25.9 | 10.146±1.463 | Sq | 0.243 | 214 |
| 471240 2011 BT15 | 21.7 | 2016-9-1 9:18 | 4×600 | 1.284 | SA93-101 (1.179) | 32.5 | 16.742±0.206 | Sr | 0.266 | 118 |
| 474163 1999 SO5 | 20.9 | 2016-11-29 3:20 | 2×1200 | 1.083 | HD6400 (1.075) | 52.2 | 19.833±0.571 | Sv | 0.309 | 158 |
| 480823 1998 YW5 | 20 | 2016-11-30 7:01 | 1×1800 | 1.224 | HD20926 (1.389) | 35.4 | 14.187±0.763 | Sv | 0.309 | 239 |
| 480922 2002 XP37 | 20.4 | 2017-1-2 1:54 | 2×750 | 1.179 | HD16640 (1.335) | 54.4 | 12.462±0.298 | Q | 0.227 | 232 |
| 482566 2012 WK4 | 21.5 | 2016-11-29 5:35 | 2×1188 | 1.211 | HD30947 (1.239) | 15.3 | 18.135±0.515 | R | 0.148 | 173 |
| 2002 RB | 20.9 | 2015-7-19 6:26 | 2×1200 | 1.048 | HD2966 (1.065) | 19.1 | -1.112±0.351 | Cb | 0.043 | 423 |
| 2005 ML13 | 22.5 | 2017-1-3 6:54 | 2×1200 | 1.574 | SA98-978 (1.528) | 27.8 | 14.360±0.583 | S | 0.211 | 91 |
| 2005 PH2 | 20.4 | 2016-3-30 7:49 | 1×840 | 1.305 | SA107-998 (1.264) | 31.1 | -0.046±1.302 | C | 0.050 | 494 |
| 2005 XT77 | 21.1 | 2015-11-6 | 1×1200 | 1.107 | HD9729 | 66.6 | 10.724±1.110 | Sq | 0.243 | 162 |
| | | 7:19 | | | (1.089) | | | | | |
| 2007 VM184 | 21 | 2016-11-29<br>0:50 | 2×240 | 1.375 | SA115-271<br>(1.218) | 29.7 | 13.557±0.264 | S | 0.211 | 183 |
| 2007 WQ3 | 21.2 | 2015-11-5<br>2:16 | 3×1200 | 1.238 | Hyades64<br>(1.439) | 18.2 | 11.935±0.411 | Sq | 0.243 | 155 |
| 2007 WU3 | 23.8 | 2015-7-20<br>1:16 | 3×1200 | 1.185 | SA110-361<br>(1.176) | 35.7 | 5.116±0.779 | X | 0.047 | 107 |
| 2008 CA6 | 20.5 | 2017-2-28<br>3:53 | 3×300 | 1.048 | HD94093<br>(1.059) | 33.9 | 12.609±0.230 | Sr | 0.266 | 205 |
| 2008 CS1 | 20.1 | 2016-8-31<br>0:09 | 2×1200 | 1.105 | HD202282<br>(1.032) | 57.8 | 13.471±0.321 | Sq | 0.243 | 257 |
| 2008 GU20 | 23.1 | 2016-4-1<br>2:12 | 3×1500 | 1.104 | SA98-978<br>(1.187) | 25.4 | 14.070±0.579 | S | 0.211 | 69 |
| 2009 CV | 24.3 | 2016-6-29<br>0:26 | 1×800 | 1.031 | Hip52311<br>(1.199) | 48.8 | 9.968±0.620 | D | 0.048 | 84 |
| 2009 DL46 | 22 | 2016-6-30<br>4:03 | 2×1200 | 1.243 | HD146997<br>(1.240) | 35.9 | 9.106±0.837 | D | 0.048 | 241 |
| 2009 EM1 | 23 | 2017-3-1<br>7:40 | 1×1800 | 1.522 | HD113171<br>(1.544) | 18.7 | 13.817±1.028 | Sq | 0.243 | 68 |
| 2009 MW | 20.8 | 30/06/16<br>01:31 | 2×600 | 1.027 | HD147034<br>(1.036) | 29.7 | 13.016±0.353 | Sq | 0.243 | 187 |
| 2011 AM24 | 20.4 | 2015-6-8<br>2:27 | 1×1350 | 1.732 | SA107-998<br>(1.265) | 49.8 | 32.067±0.833 | D? | 0.2? | 247? |
| 2011 OL5 | 20.2 | 2015-6-8<br>7:04 | 2×750 | 1.493 | SA110-361<br>(1.274) | 36.7 | 11.399±0.435 | Xe | 0.136 | 329 |
| 2012 DR32 | 24.8 | 2017-2-28<br>6:02 | 2×400 | 1.619 | HD88749<br>(1.188) | 26.7 | 4.281±0.301 | Xc | 0.129 | 41 |
| 2012 NP | 21.3 | 2015-7-20<br>2:17 | 2×900 | 1.356 | HD111244<br>(1.060) | 40.7 | 30.543±0.527 | A | 0.191 | 167 |
| 2012 PG6 | 20.3 | 2015-7-19<br>7:10 | 2×900 | 1.05 | HD2966<br>(1.065) | 14.1 | 5.245±0.871 | X | 0.047 | 534 |
| 2012 RS16 | 21.4 | 2015-7-19<br>23:45 | 2×1200 | 1.329 | SA107-998<br>(1.233) | 35 | 24.760±1.004 | V | 0.362 | 116 |
| 2012 XA133 | 21 | 2015-12-15<br>1:34 | 2×900 | 1.228 | Hip9329<br>(1.043) | 44 | 6.494±1.440 | Q | 0.227 | 176 |
| 2013 YE38 | 20.1 | 2016-11-30<br>8:19 | 1×900 | 1.098 | HD20926<br>(1.062) | 57.3 | 10.396±0.309 | D | 0.048 | 579 |
| 2014 GF50 | 20.7 | 2016-8-31<br>2:12 | 3×1200 | 1.059 | SA110-361<br>(1.149) | 34.6 | 18.374±0.437 | Sq | 0.243 | 195 |
| 2014 OE338 | 21 | 2015-7-20<br>4:45 | 2×1200 | 1.272 | SA110-361<br>(1.176) | 22.7 | 7.503±1.108 | Xc | 0.129 | 233 |
| 2014 QK362 | 21.6 | 2015-7-19<br>4:22 | 3×900 | 1.008 | SA112-1333<br>(1.366) | 1.6 | 1.217±0.524 | Cb | 0.043 | 307 |
| 2014 TF17 | 20.8 | 2015-4-14<br>8:33 | 2×600 | 1.235 | SA102-1081<br>(1.202) | 32.8 | 4.129±1.075 | X | 0.047 | 424 |
| 2014 WP365 | 20.2 | 2015-4-14<br>2:59 | 1×300 | 1.089 | SA102-1081<br>(1.202) | 48.9 | 11.783±1.244 | S | 0.211 | 264 |
| 2014 YC | 22.2 | 2017-1-2<br>3:48 | 1×1333 | 1.087 | SA102-1081<br>(1.142) | 32.1 | 6.973±0.824 | Xc | 0.129 | 134 |
| 2014 YS34 | 20.8 | 2015-6-9<br>7:14 | 2×1125 | 1.249 | SA102-1081<br>(1.556) | 20.1 | 24.958±0.978 | A | 0.191 | 210 |
| 2015 AY245 | 21.1 | 2015-7-20<br>3:27 | 3×1200 | 1.369 | HD154424<br>(1.234) | 32.4 | 6.690±0.674 | T | 0.042 | 391 |
| 2015 BY310 | 21.7 | 2015-6-8<br>9:44 | 2×1200 | 1.11 | SA102-1081<br>(1.149) | 44.5 | 12.418±0.954 | Sr | 0.266 | 118 |
| 2015 DR215 | 20.5 | 2016-4-1<br>0:29 | 2×900 | 1.624 | SA107-998<br>(1.353) | 92.1 | 17.527±1.265 | Sr | 0.266 | 205 |
| 2015 FD134 | 20.4 | 2015-6-9<br>2:06 | 1×1200 | 1.189 | SA102-1081<br>(1.556) | 48.6 | 17.291±0.904 | Sv | 0.309 | 199 |
| 2015 GF | 20.9 | 2015-6-8<br>8:28 | 2×900 | 1.172 | SA115-271<br>(1.185) | 27.6 | 16.096±0.557 | Sr | 0.266 | 170 |
| 2015 HA1 | 21.2 | 2015-6-9 1:09 | 1×1200 | 1.441 | SA102-1081 (1.556) | 82.3 | 13.890±0.533 | L | 0.149 | 198 |
| 2015 HB117 | 23.6 | 2015-6-8 4:24 | 2×1200 | 1.359 | SA102-1081 (1.149) | 33 | 26.095±0.979 | A | 0.191 | 58 |
| 2015 HM10 | 23.6 | 2015-7-20 8:10 | 2×1200 | 1.26 | HD111244 (1.060) | 63.4 | 18.302±1.068 | R | 0.148 | 66 |
| 2015 HP43 | 21.1 | 2015-6-8 3:00 | 1×900 | 1.047 | SA115-271 (1.185) | 23 | 8.330±0.556 | Q | 0.227 | 168 |
| 2015 JD1 | 20.7 | 2015-11-7 8:51 | 3×400 | 1.114 | HD625558 (1.102) | 63.7 | 16.824±1.502 | L | 0.149 | 249 |
| 2015 JJ2 | 21.9 | 2015-6-9 4:45 | 1×900 | 1.111 | SA110-361 (1.151) | 16.8 | 2.919±0.757 | C | 0.050 | 248 |
| 2015 JY1 | 20.8 | 2015-6-8 0:16 | 2×750 | 1.003 | SA102-1081 (1.149) | 47 | 15.732±0.289 | R | 0.148 | 239 |
| 2015 KS121 | 22.8 | 2015-6-9 5:50 | 1×600 | 1.139 | SA107-998 (1.149) | 10.9 | 3.075±1.473 | X | 0.047 | 169 |
| 2015 KU121 | 22.9 | 2015-6-7 23:25 | 1×150 | 1.117 | SA102-1081 (1.149) | 80.1 | 12.465±0.615 | Q | 0.227 | 73 |
| 2015 LH | 27.3 | 2015-6-9 6:30 | 1×180 | 1.426 | SA102-1081 (1.556) | 43.5 | 37.935±1.700 | A | 0.191 | 11 |
| 2015 LH14 | 20.1 | 2015-7-19 0:35 | 1×900 | 1.383 | SA107-998 (1.269) | 41 | 11.873±0.454 | Xc | 0.129 | 353 |
| 2015 LN21 | 23 | 2015-7-1 9:22 | 2×1200 | 1.425 | SA115-271 (1.290) | 54.1 | 19.597±0.680 | D | 0.048 | 152 |
| 2015 LU24 | 20.4 | 2015-7-20 6:35 | 2×1200 | 1.391 | SA110-361 (1.176) | 57.2 | 19.405±1.047 | Sa | 0.367 | 182 |
| 2015 MN44 | 22.5 | 2015-7-19 5:26 | 3×1200 | 1.042 | HD2966 (1.065) | 9.9 | 7.278±0.518 | Q | 0.227 | 88 |
| 2015 QQ3 | 21.3 | 2015-11-5 4:34 | 3×1200 | 1.058 | HD9729 (1.052) | 30.7 | 8.524±0.519 | Sq | 0.243 | 148 |
| 2015 RD37 | 20.1 | 2015-11-5 3:19 | 2×600 | 1.032 | HD11123 (1.021) | 25.1 | 15.448±0.452 | V | 0.362 | 211 |
| 2015 RG36 | 20.3 | 2015-11-6 0:48 | 2×1100 | 1.046 | HD9729 (1.089) | 56.6 | 11.329±0.430 | S | 0.211 | 252 |
| 2015 TA | 21.6 | 2015-11-7 1:37 | 2×1800 | 1.046 | HD7186 (1.040) | 50.6 | 16.424±0.901 | S | 0.211 | 138 |
| 2015 TA25 | 20.1 | 2015-11-7 3:28 | 2×600 | 1.655 | HD9729 (1.452) | 11.7 | 11.422±0.250 | S | 0.211 | 276 |
| 2015 TB179 | 20.6 | 2015-11-5 7:34 | 3×1200 | 1.203 | HD11123 (1.021) | 27.1 | 25.344±0.713 | A | 0.191 | 231 |
| 2015 TG238 | 22.8 | 2015-11-7 2:49 | 2×1200 | 1.181 | HD9729 (1.452) | 15.4 | 2.841±0.982 | X | 0.047 | 169 |
| 2015 TJ1 | 22.7 | 2016-3-31 2:53 | 1×1200 | 1.077 | SA102-1081 (1.288) | 37.8 | 1.142±0.833 | Cb | 0.043 | 185 |
| 2015 TK238 | 21.9 | 2015-11-6 4:44 | 1×1200 | 1.168 | SA112-1333 (1.260) | 17.8 | 7.063±0.611 | Q | 0.227 | 116 |
| 2015 TL143 | 23.3 | 2015-11-6 3:15 | 2×900 | 1.051 | SA112-1333 (1.260) | 55.8 | 13.618±0.360 | Sv | 0.309 | 52 |
| 2015 TM143 | 23.6 | 2015-11-6 2:29 | 2×1200 | 1.13 | HD3011 (1.120) | 69.2 | -0.348±0.596 | Cb | 0.043 | 122 |
| 2015 TW144 | 20.7 | 2015-11-5 5:46 | 2×900 | 1.285 | Hyades64 (1.439) | 21.5 | 27.093±0.531 | A | 0.191 | 220 |
| 2015 TY144 | 21.4 | 2015-11-6 7:48 | 1×1200 | 1.132 | HD9729 (1.089) | 33.7 | 11.559±0.715 | Q | 0.227 | 146 |
| 2015 TZ237 | 24.3 | 2015-11-7 4:13 | 2×1200 | 1.571 | HD9729 (1.452) | 4.9 | 6.030±0.577 | X | 0.047 | 85 |
| 2015 UC | 24.8 | 2015-11-7 5:02 | 2×1200 | 1.183 | HD9729 (1.452) | 13.8 | 11.450±0.326 | Sq | 0.243 | 30 |
| 2015 UJ51 | 21.5 | 2015-11-7 5:53 | 2×1200 | 1.244 | HD625558 (1.102) | 12.6 | 6.906±0.503 | Q | 0.227 | 140 |
| 2015 UK52 | 20.1 | 2015-12-14 | 1×900 | 1.194 | SA93-101 | 77.2 | 13.185±0.451 | Sr | 0.266 | 246 |
| | | 1:29 | | | (1.152) | | | | | |
| 2015 UT52 | 20.9 | 2015-12-14 3:30 | 1×600 | 1.453 | Hyades64 (1.455) | 7.2 | 4.536±0.333 | X | 0.047 | 405 |
| 2015 WG9 | 20.3 | 2015-12-15 3:16 | 3×900 | 1.139 | HD2966 (1.101) | 35.8 | 6.396±1.029 | Sq | 0.243 | 235 |
| 2015 XE | 24.6 | 2015-12-14 5:48 | 2×900 | 1.368 | HD2966 (1.079) | 17.3 | 7.894±0.367 | K | 0.130 | 44 |
| 2015 XK1 | 20 | 2015-12-14 4:36 | 2×600 | 1.197 | HD2966 (1.079) | 27.8 | 9.671±0.326 | Q | 0.227 | 279 |
| 2016 CB30 | 23.6 | 2016-3-30 4:29 | 3×800 | 1.162 | SA102-1081 (1.297) | 11.9 | 0.105±0.447 | Cb | 0.043 | 122 |
| 2016 CZ135 | 20.7 | 2016-11-30 2:00 | 1×1500 | 1.113 | HD4941*1.09 7 | 27.6 | 13.672±0.638 | Sv | 0.309 | 173 |
| 2016 EC157 | 23.5 | 2016-4-1 3:28 | 2×1800 | 1.163 | SA98-978 (1.187) | 15.6 | 11.396±0.738 | S | 0.211 | 58 |
| 2016 EK27 | 22.3 | 2016-4-1 5:21 | 1×750 | 1.01 | HD106649 (1.010) | 31 | 11.306±0.462 | S | 0.211 | 100 |
| 2016 LM1 | 20 | 2016-7-1 9:38 | 2×1200 | 1.019 | HD3011 (1.009) | 61.2 | 13.535±0.697 | L | 0.149 | 344 |
| 2016 LU10 | 20.9 | 2016-9-1 8:02 | 3×1200 | 1.159 | HD202282 (1.032) | 32 | 2.471±0.313 | C | 0.050 | 393 |
| 2016 LZ10 | 20 | 2016-7-1 4:37 | 3×900 | 1.175 | HD125910 (1.071) | 21 | 3.027±0.466 | B | 0.120 | 384 |
| 2016 MF1 | 20.9 | 2016-8-30 2:50 | 1×1200 | 1.074 | HD202282 (1.038) | 9.4 | 6.299±0.507 | O | 0.339 | 151 |
| 2016 OJ1 | 21.7 | 2016-8-29 7:42 | 1×1200 | 1.019 | SA93-101 (1.181) | 30.6 | 13.701±1.433 | Sv | 0.309 | 109 |
| 2016 OK1 | 20 | 2017-1-3 1:20 | 2×944 | 1.121 | HD11675 (1.102) | 60.1 | 13.784±0.405 | Q | 0.227 | 279 |
| 2016 PA8 | 21.9 | 2016-9-1 6:36 | 3×1200 | 1.207 | SA115-271 (1.155) | 3.9 | 9.712±0.496 | Sq | 0.243 | 112 |
| 2016 PB8 | 22.5 | 2016-8-31 3:49 | 3×1200 | 1.021 | HD202282 (1.032) | 10 | 2.064±0.302 | C | 0.050 | 188 |
| 2016 PJ66 | 22.2 | 2016-9-1 5:09 | 2×1200 | 1.054 | HD202282 (1.032) | 8.6 | 14.037±0.545 | Sr | 0.266 | 94 |
| 2016 PN38 | 21.2 | 2016-8-30 0:21 | 3×1200 | 1.073 | HD202282 (1.038) | 30 | 6.050±0.517 | X | 0.047 | 353 |
| 2016 PU | 21.7 | 2016-8-30 8:19 | 2×1200 | 1.093 | HD202282 (1.038) | 33.3 | 13.27*0.327 | V | 0.362 | 101 |
| 2016 QF44 | 21.4 | 2016-11-30 3:03 | 2×1500 | 1.419 | Hyades64 (1.446) | 6.8 | 15.31*0.589 | Sv | 0.309 | 125 |
| 2016 TD18 | 23.6 | 2016-11-29 6:36 | 1×900 | 1.047 | HD30947 (1.239) | 55.1 | 10.412*1.255 | Sq | 0.243 | 51 |
| 2016 UU80 | 21.1 | 2016-11-30 6:18 | 1×1500 | 1.304 | Hyades64 (1.446) | 32.4 | 9.638±0.610 | Sv | 0.309 | 144 |
| 2016 VY5 | 20.2 | 2017-1-3 2:00 | 3×630 | 1.198 | HD11675 (1.102) | 57.9 | 17.948±0.300 | S | 0.211 | 264 |
| 2016 WB8 | 25.9 | 2016-11-30 4:26 | 1×1200 | 1.151 | HD20926 (1.062) | 29.5 | 13.013±0.613 | L | 0.149 | 23 |
| 2016 WG7 | 26.1 | 2016-11-29 4:26 | 1×300 | 1.455 | Hyades64 (1.439) | 13.9 | 14.873±0.488 | Sq | 0.243 | 16 |
| 2016 WJ1 | 21.3 | 2016-11-30 4:01 | 1×450 | 1.378 | Hyades64 (1.446) | 9.2 | 10.898±0.311 | Sq | 0.243 | 148 |
| 2016 WL7 | 24.3 | 2016-11-29 2:06 | 2×900 | 1.111 | HD6400 (1.075) | 64.3 | 12.045±0.268 | D | 0.048 | 84 |
| 2016 WQ | 25.5 | 2016-11-29 1:15 | 1×450 | 1.077 | SA115-271 (1.218) | 47.2 | 9.153±1.014 | Q | 0.227 | 22 |
| 2016 WZ8 | 28.4 | 2016-11-30 4:59 | 1×600 | 1.458 | HD20926 (1.389) | 6.4 | 10.155±0.689 | D | 0.048 | 13 |
| 2016 YF | 25.4 | 2017-1-2 6:38 | 2×1800 | 1.129 | SA102-1081 (1.142) | 57.5 | 12.254±0.576 | S | 0.211 | 24 |
| 2016 YM1 | 22.2 | 2017-3-1 2:36 | 2×1500 | 1.225 | SA98-978 (1.180) | 10.7 | 4.530±0.524 | Cg | 0.063 | 192 |
| 2017 BL30 | 23.3 | 2017-2-28 1:03 | 1×900 | 1.233 | SA98-978 (1.145) | 63.3 | 14.158±1.287 | S | 0.211 | 63 |
| 2017 DA36 | 25.1 | 2017-2-28 2:19 | 2×800 | 1.165 | SA102-1081 (1.258) | 22.9 | 4.883±0.744 | Xk | 0.095 | 41 |
| 2017 DA38 | 25 | 2017-3-1 5:52 | 2×750 | 1.415 | HD113171 (1.544) | 29 | 15.148±0.296 | Sq | 0.243 | 27 |
| 2017 DC36 | 22.1 | 2017-2-28 3:17 | 2×400 | 1.087 | SA102-1081 (1.258) | 20.4 | 8.361±0.163 | S | 0.211 | 110 |
| 2017 DL34 | 25.9 | 2017-2-28 5:22 | 2×1600 | 1.147 | HD91640 (1.083) | 8.9 | 9.057±0.769 | D | 0.048 | 40 |
| 2017 DR15 | 20.9 | 2017-2-28 4:28 | 2×900 | 1.386 | HD94093 (1.059) | 5.8 | 13.323±0.443 | S | 0.211 | 191 |
| 2017 DV15 | 25.1 | 2017-2-28 7:40 | 1×1500 | 1.239 | SA102-1081 (1.258) | 50.2 | 16.559±0.548 | Sv | 0.309 | 23 |

**Notes: The equivalent diameter of each object is computed by taking into account the mean albedo of its assigned taxon (as from Mainzer et al., 2011a), except when an albedo measurement is available in the literature (reported values from: (1) Mueller et al., 2011, (2) Trilling et al., 2010, (3) Mainzer et al., 2014, (4) Mainzer et al., 2011b).**

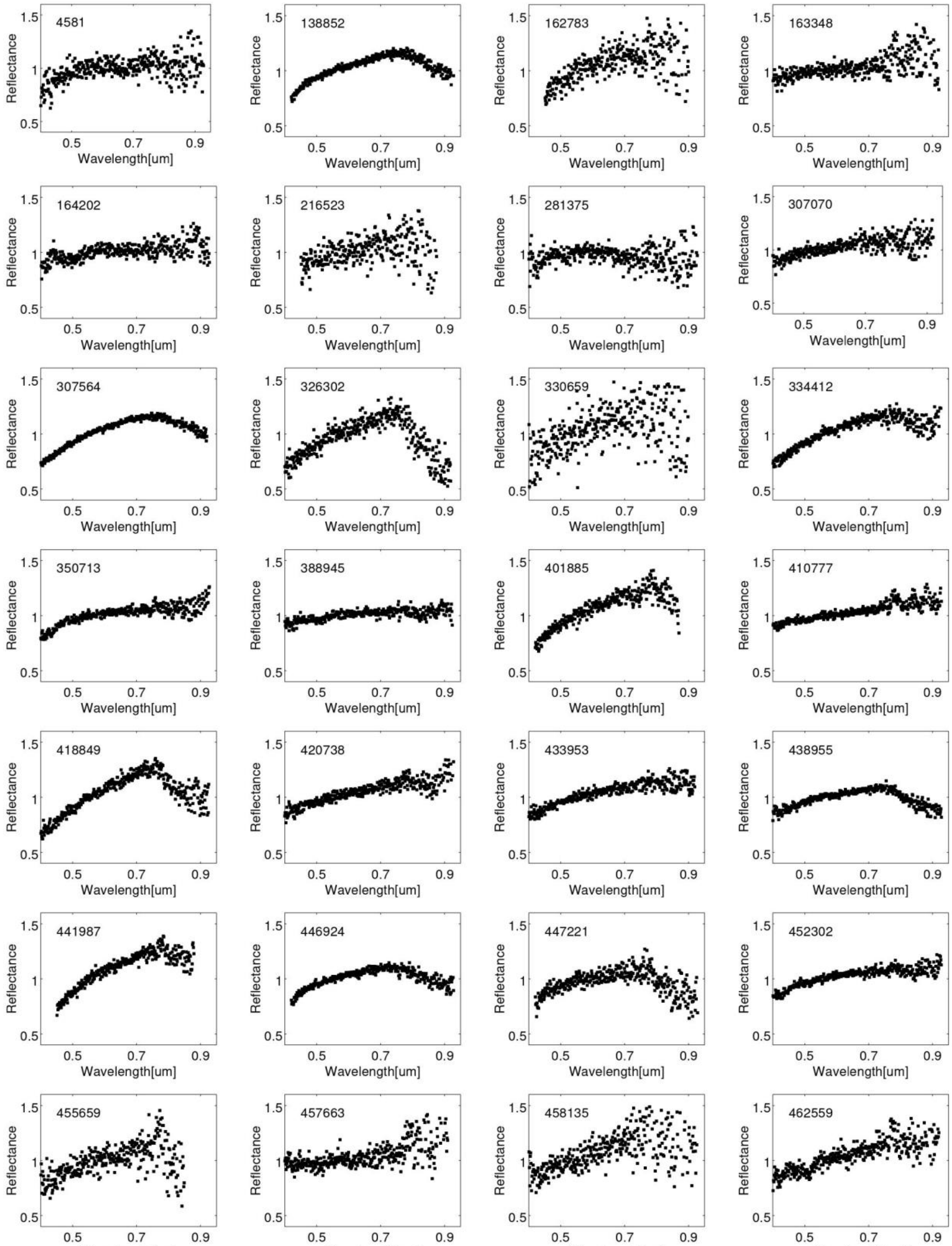

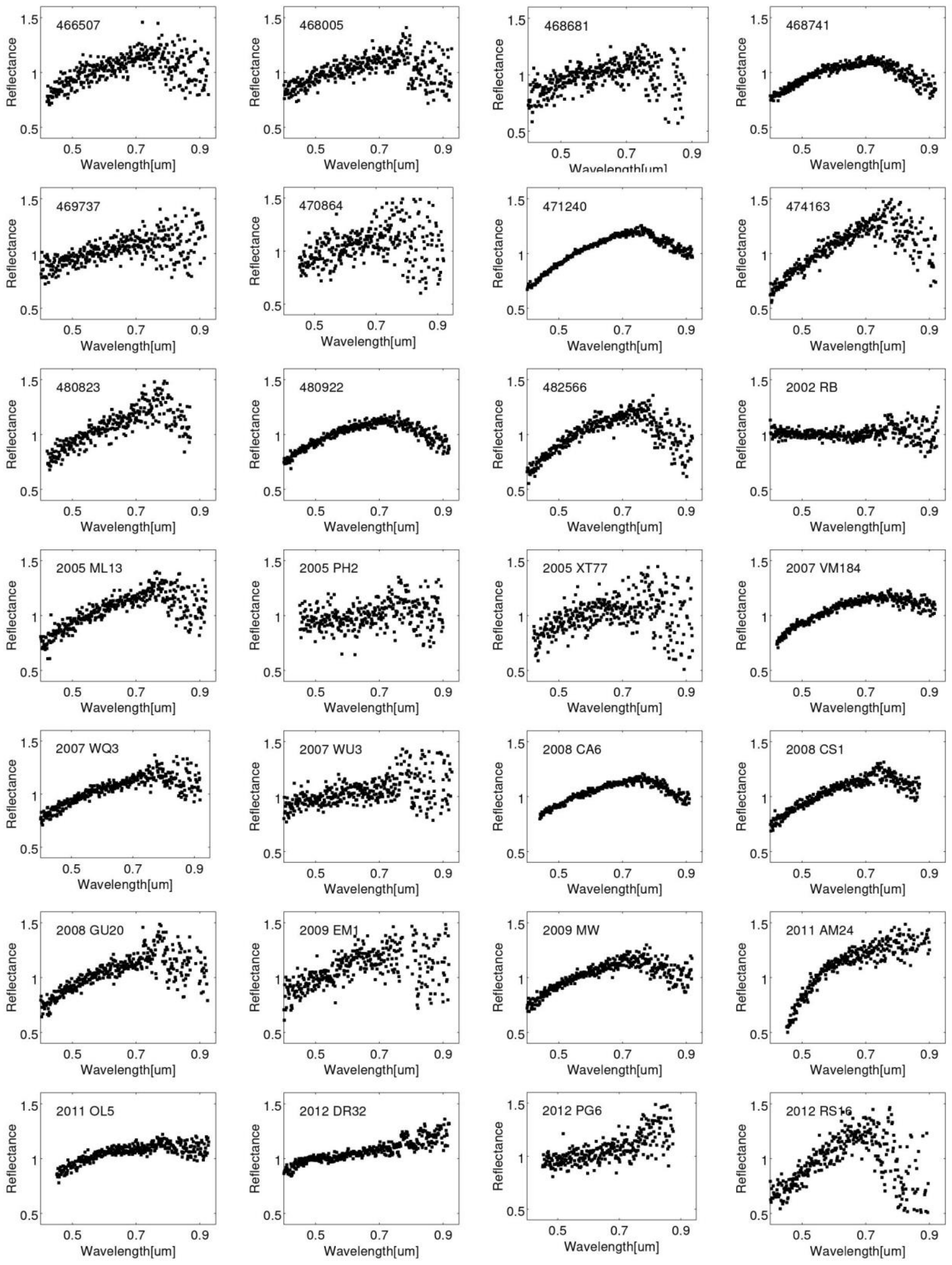

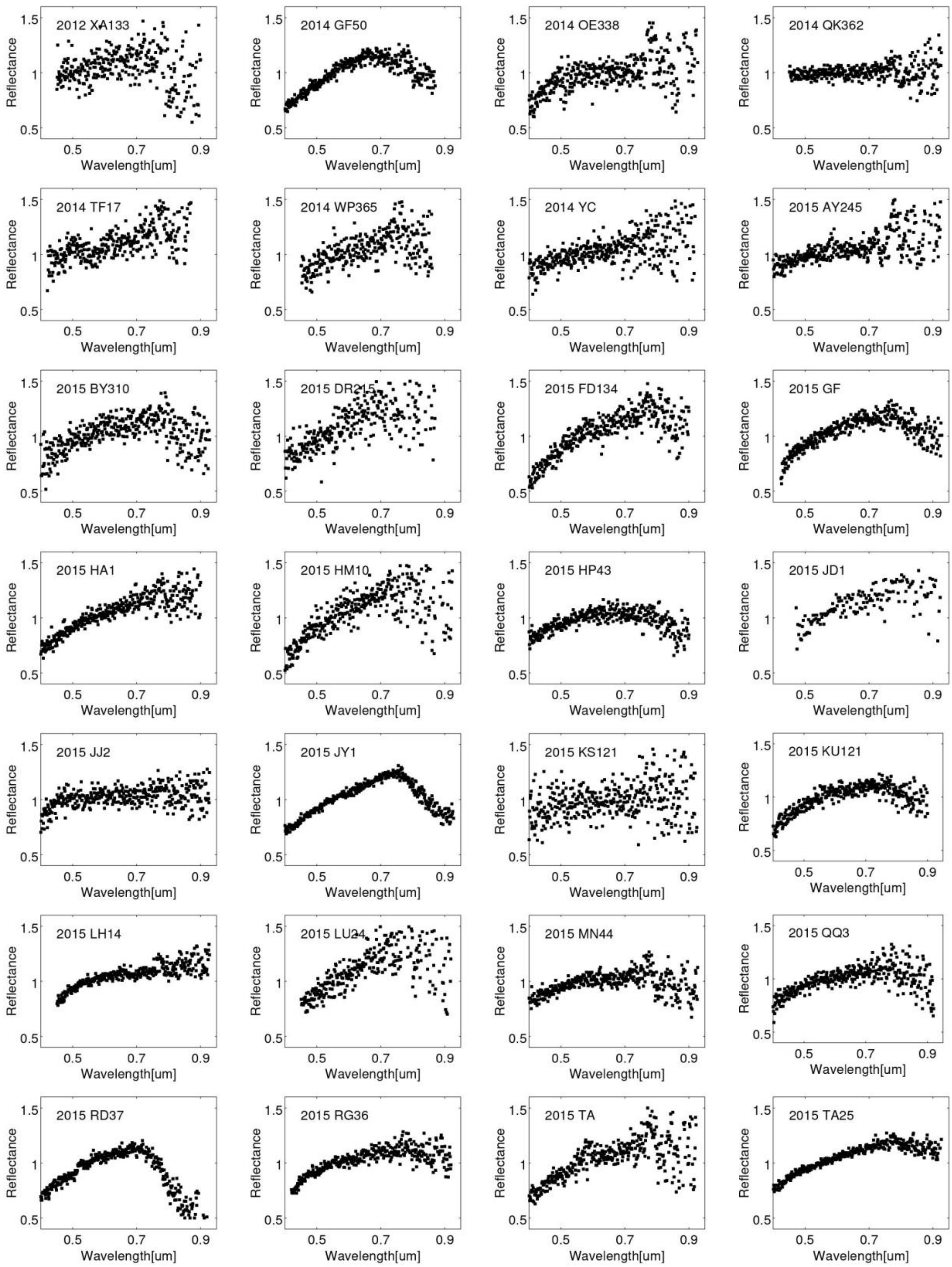

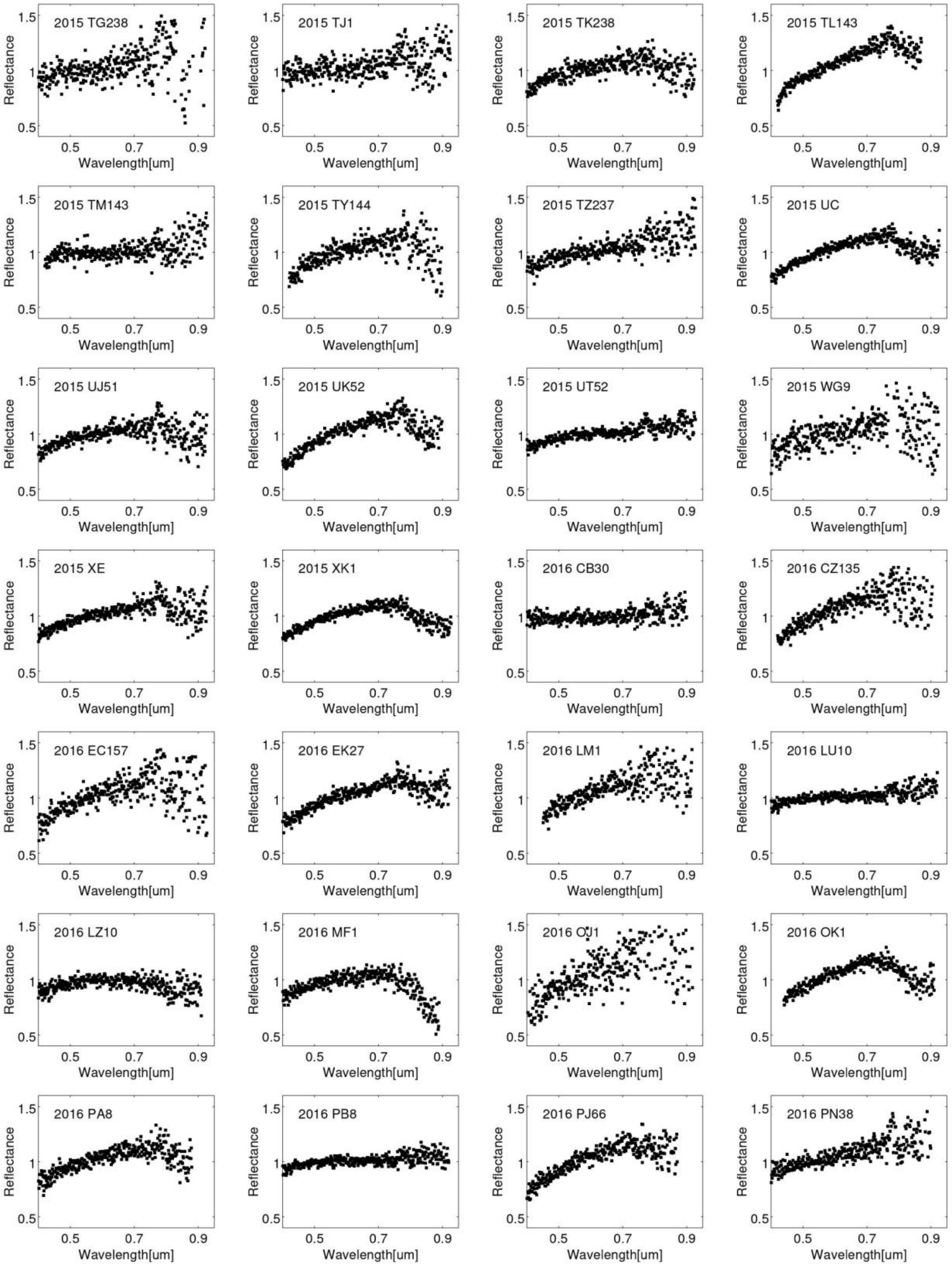

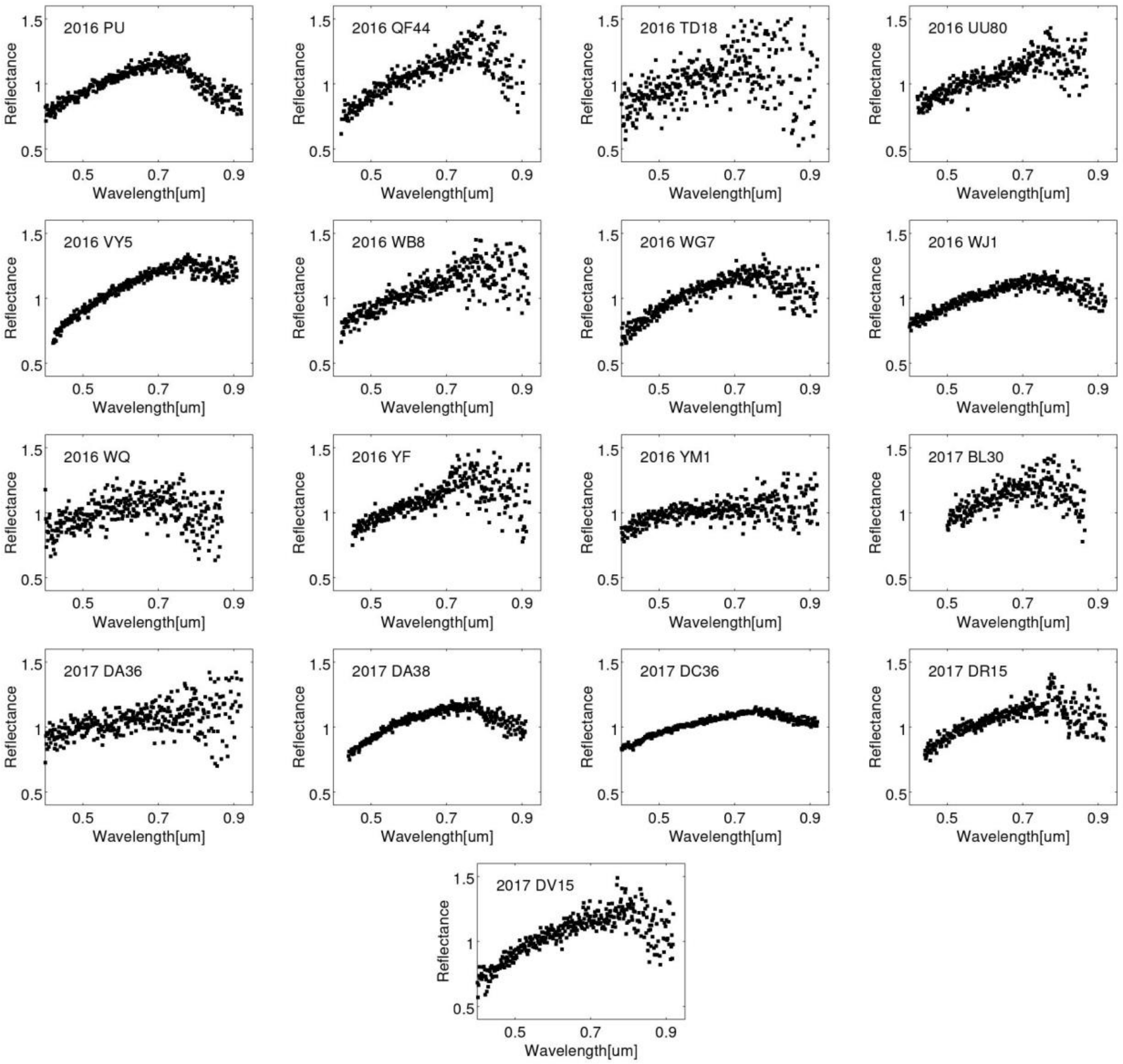

**Figure 2: Obtained reflectance spectra (normalized at 0.55 μm) for 129 NEAs (spectra of further 18 targets are presented in parallel papers, cf. text in Section 2).**

# 3 Results

## 3.1 Taxonomic distribution

The taxonomic distribution of the observed NEAs is shown in Fig. 3. We report separate statistics for objects in the size bins 0-100 m and 100-300 m. Taking into account the uncertainties due to the availability of visible spectra only, to increase the significance of our analysis we group together objects belonging to the S-, C- and X-complexes. Moreover, we put in the X-complex the only target we classified as a T-type (with the objective of keeping well separated the primitive D-types from other featureless spectra in the X-complex). The distribution we found from our observations is dominated by the S-complex at all sizes, in agreement with previous results for larger NEAs. Such dominance is the consequence of both observational biases (albedo and phase angle effects) and preferential transport mechanisms from the inner asteroid main belt where S-type asteroids are dominant (e.g., Binzel et al., 2004). A proper debiasing of our observational results is beyond the scope of this paper, as it would be very much complicated by the still low statistics available for the small NEAs. We stress however that due to our constraint in the target selection concerning the absolute magnitude (only objects with H ≥ 20 were considered), all of the larger objects (> 300 m) in our sample are basically low-albedo ones. Conversely, one should remind of a bias against the discovery/observation of small low-albedo objects, which makes them very probably underrepresented in the 100-300 m and (especially) in the 0-100 m size ranges.

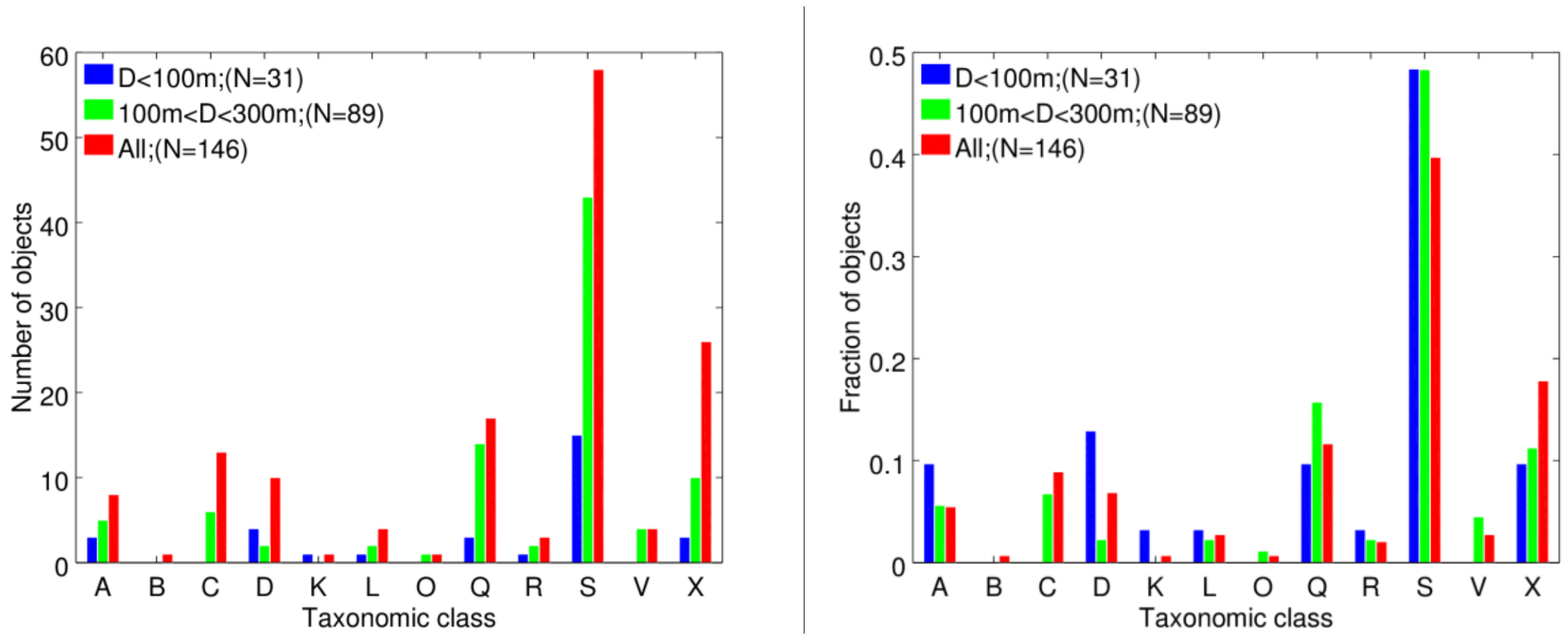

**Figure 3: Taxonomic type distribution of the observed NEAs (left: absolute numbers; right: relative percentages). The distributions for objects smaller than 100 m and for objects in the 100-300 m size bin are shown separately.**

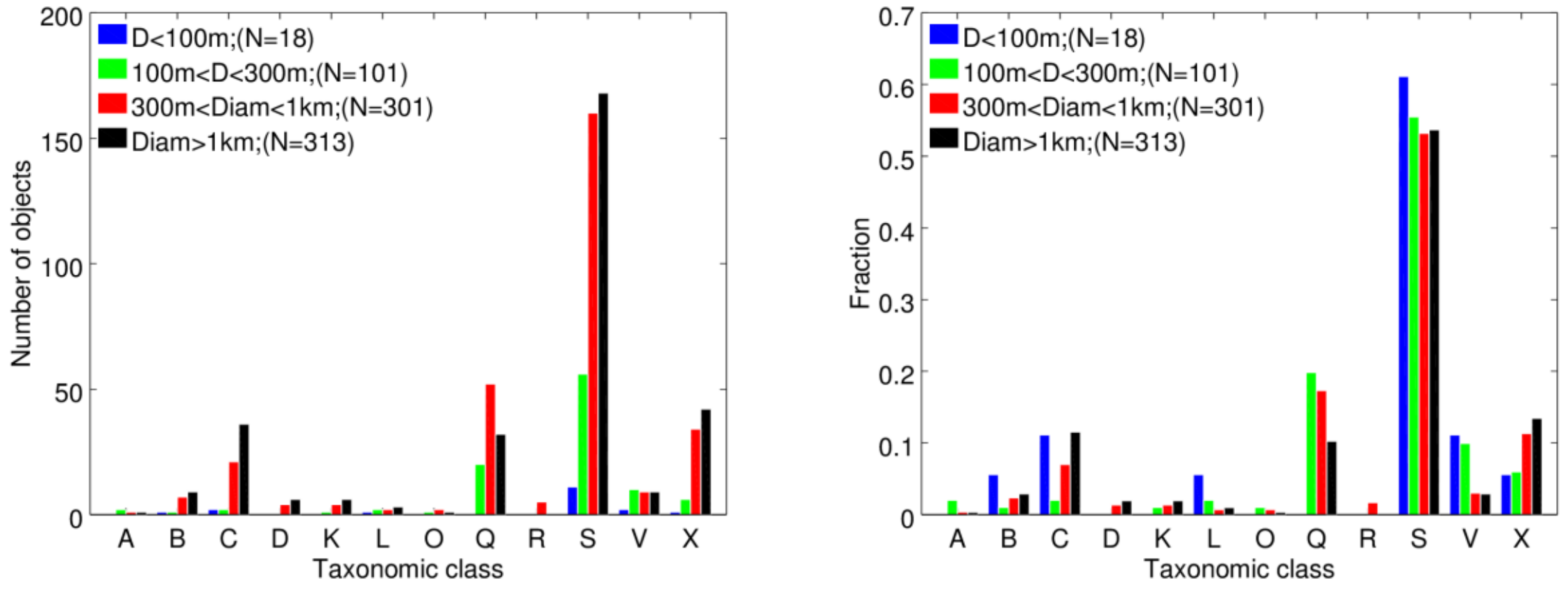

**Figure 4: Taxonomic type distribution of NEAs in the EARN database (left: absolute numbers; right: relative percentages). The distributions for objects with size in the ranges <100 m, 100-300 m, 300-1000 m and >1 km, are shown separately.**

**Table 2: Statistical comparison between taxonomic distributions of NEAs in our sample and in the literature.**

| Taxon | A | B | C | D | K | L | O | Q | R | S | V | X |
|---|---|---|---|---|---|---|---|---|---|---|---|---|
| **Our sample (%)** | 5.48 (8/146) | 0.68 (1/146) | 8.90 (13/146) | 6.85 (10/146) | 0.68 (1/146) | 2.74 (4/146) | 0.68 (1/146) | 11.64 (17/146) | 2.05 (3/146) | 39.73 (58/146) | 2.74 (4/146) | 17.81 (26/146) |
| **146-unit random samples from EARN database (%) [±1σ]** | 0.47± 0.56 | 2.45± 1.27 | 8.38± 2.26 | 1.38± 0.97 | 1.49± 0.99 | 1.09± 0.87 | 0.54± 0.61 | 14.23± 2.86 | 0.68± 0.67 | 53.94± 4.11 | 4.10± 1.62 | 11.25± 2.61 |
| **Deviation (σ)** | +8.9 | -1.4 | +0.2 | +5.6 | -0.8 | +1.9 | +0.2 | -0.9 | +2.0 | -3.5 | -0.8 | +2.5 |

For comparison, Fig. 4 shows the distribution of taxonomic types in the literature, retrieved from the European Asteroid Research Node (EARN)[3] database of NEA physical properties. Taxonomic information for a total of 733 objects is reported, for different size bins (0-100 m, 100-300 m, 300-1000 m, >1 km). We use the same grouping criteria as in Fig. 3.

While the size distributions of our targets and of NEAs in the EARN database partly overlap, we can in first approximation consider them as representing "smaller" and "larger" asteroids, respectively. To compare statistically these two datasets in terms of taxonomic distribution, we used the following approach: (i) we randomly selected 146 objects from the EARN sample (i.e. the same number of NEAs characterized in our survey); (ii) we checked the taxonomic distribution of such random subsample; (iii) we repeated 10000 times steps (i) and (ii), and recorded the obtained mean and standard deviation for the population of each taxon. The results of such analysis are reported in Table 2. The most striking difference emerging when comparing our sample (i.e., "smaller" NEAs) with data reported in the literature (i.e., "larger" NEAs) concerns the overabundance of A- and D-types within the "small" NEA population, with deviations of about 5.6σ and 8.9σ with respect to the expected population of these taxa if the taxonomic distribution of NEAs in the EARN database could be assumed as the reference distribution. Being based on visible data only, our findings about A-types have to reckon with some uncertainties due to possible misassignments with respect to other taxa presenting the 1-μm silicate band (we incidentally stress that we find the S-complex asteroids underrepresented in our sample, with respect to larger NEAs, with a deviation of -3.5σ). However, the results seem statistically very robust, also in consideration of the overall good quality of our spectra. The abundance of D-types (4 of which with diameter < 100 m) among our targets looks even more glittering, considering the low albedo of these objects that should disadvantage their detection at the smallest size ranges. Noteworthy, we also outline a relatively high abundance in our sample (+2.5σ) of featureless X-type NEAs, part of which could be of carbonaceous, primitive nature (see Fornasier et al., 2011, for a review on X-type asteroids). The populations of the rest of the taxa do not differ by more than 2σ with respect to the expected values based on the literature data. The possible implications of an overabundance of organic-rich D-types and olivine-rich A-types among the small NEA population are further discussed in Sec. 4.

[3] http://earn.dlr.de; retrieved on 13 August 2017.

## 3.2 Phase reddening

Phase reddening (i.e., the increasing of the spectral slope with increasing solar phase angle $\alpha$ at the moment of the observation) has been evidenced to affect many bodies of the solar system, from the Moon to Mercury, from Uranus' moons to asteroids (e.g., Schröder et al., 2014, and references therein).

Ideally, to study the phase reddening one should observe each object at different phase angles, and obtain particular coefficients for each asteroid. Of course this would be very much time-demanding if a large population of bodies has to be investigated. However, our observations of NEAs have been performed over a wide range of solar phase angles (~2-94 degrees), allowing us to constrain the "average" phase reddening effect on the visible spectral slope (0.44-0.65 μm; cf. Table 1) for the different taxonomic types. We warn however the reader of the limitations of our approach, as in this way we cannot separate the phase reddening due to the observing geometry from other factors affecting the spectral properties of each asteroid, such as space weathering, particle size effects, etcetera.

The average values of the reddening coefficient $\gamma$ (e.g., Luu & Jewitt, 1990), defined as the linear fit of the spectral slope vs. phase angle dependence, are given in Table 3 for the most populated compositional groups in our sample (together with the number of objects in each taxon and the measured range of phase angles; again, we used the Octave's *polyfit* routine).

The phase reddening seems to strongly affect moderate-to-high albedo spectral types. The largest phase reddening is seen for olivine-rich A-types (and, to a less extent, Q-types). Asteroids belonging to the S-complex barely show phase reddening, with a large data scatter most probably due to a lower level of composition homogeneity within this taxon (Fig. 5).

Conversely to silicate-rich asteroids, NEAs in the C-complex do not show phase reddening (in particular, disregarding the only point at $\alpha$ > 50 degrees one would obtain $\gamma$=-0.010±0.013). The visible spectral slopes of the D-types also seem to not increase with the phase angle (disregarding the two reddest D-types, whose spectral slopes are considerably higher than others reported for this class, one would obtain $\gamma$=0.053±0.027). In both cases, the data scatter is rather large. The X-complex also shows large dispersions in spectral slopes (probably because of the coexistence in this taxon of objects of either carbonaceous, silicaceous, enstatitic, or metallic nature), although on average a reddening trend could be present (Fig. 6).

**Table 3: Reddening coefficient γ for different spectral types.**

| Taxon | Phase angle range (deg) | Number of objects | γ (%/100 nm/deg) [0.44-0.65 μm] |
|---|---|---|---|
| A | 20-51 | 8 | 0.492±0.072 |
| Q | 10-80 | 17 | 0.074±0.018 |
| S | 4-92 | 58 | 0.013±0.009 |
| X | 5-56 | 26 | 0.074±0.021 |
| D | 6-64 | 10 | -0.024±0.027 |
| C | 2-69 | 13 | -0.032±0.030 |

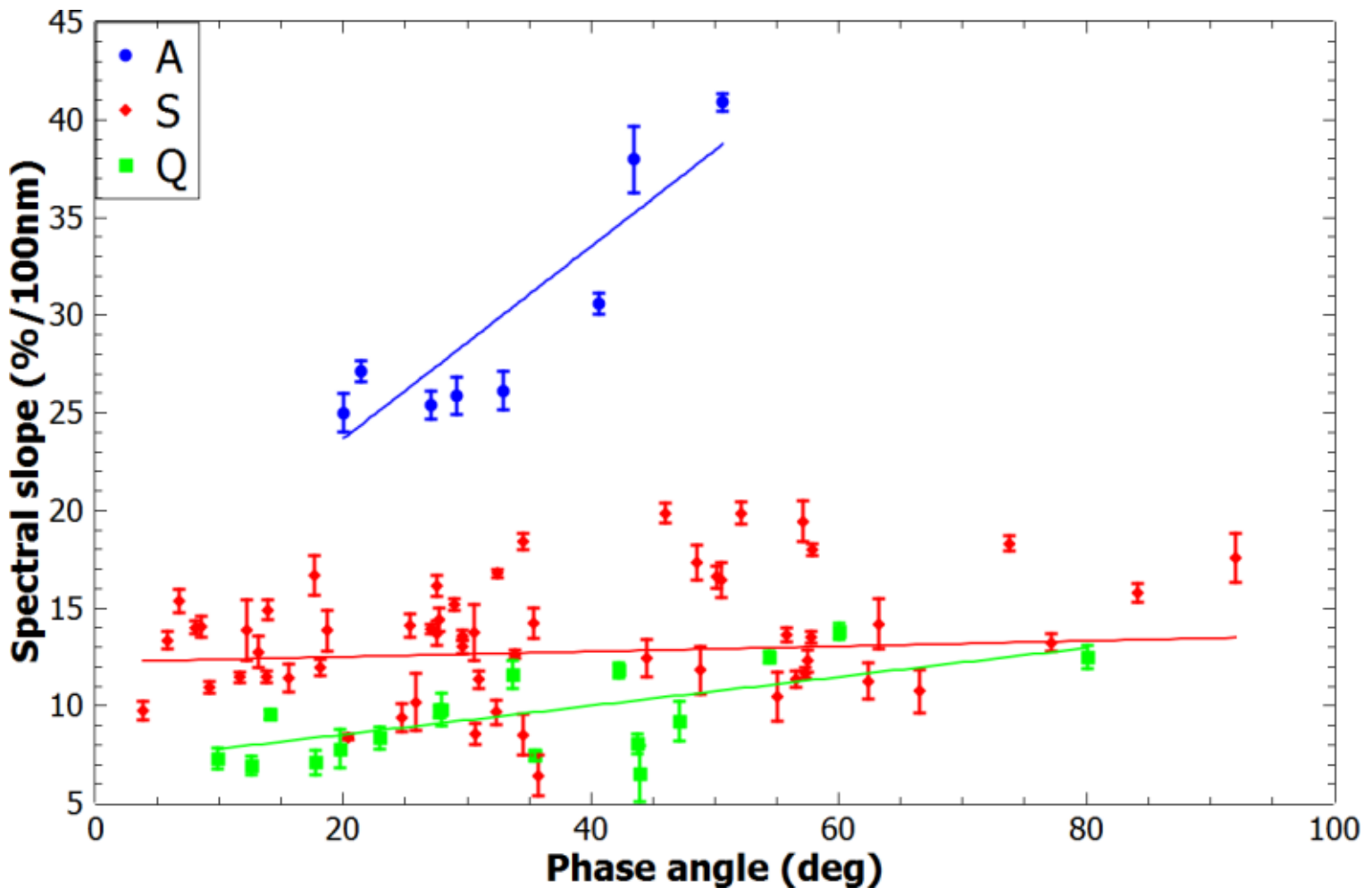

**Fig. 5: Spectral slope (0.44-0.65 μm) vs. phase angle for A-, S- and Q-types (blue dots, red diamonds and green squares, respectively). Linear fits are also reported with continuous lines for the three compositional groups (cf. Table 3).**

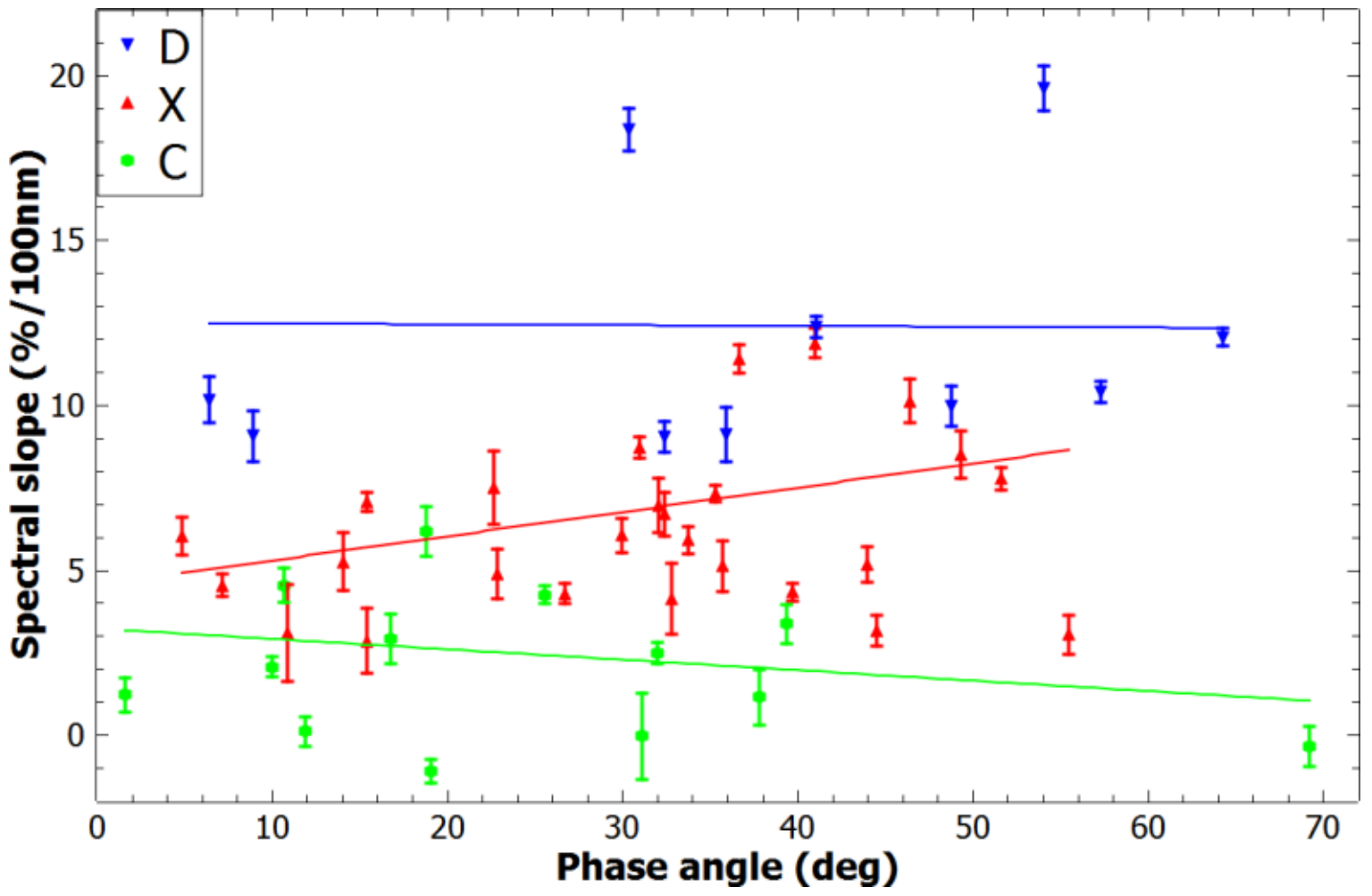

**Fig. 6: Spectral slope (0.44-0.65 μm) vs. phase angle for D-, X- and C-types (blue down-triangles, red up-triangles and green dots, respectively). Linear fits are also reported with continuous lines for the three compositional groups (cf. Table 3).**

## 4 Discussion and conclusions

Despite of its intrinsic limitations (visible data only; still low statistics), our visible spectroscopic survey of the “small” (H≥20) near-Earth asteroids reveals some peculiar characteristics of the distribution of taxonomic types within such population, compared to larger bodies.

First of all, we note the relative abundance of olivine-rich A-types (5-10%). Such bodies are very rare among larger NEAs and in the asteroid main belt (where objects of just a few hundred metres in size are not observable). Here we remind the reader of the long-lasting “missing olivine problem” (e.g., Scott et al., 2015, and references therein): decay of $^{26}$Al in the early solar system produced a differentiation of the largest accreting planetesimals. Later mutual collisions shattered these early-formed differentiated bodies to produce a much bigger number of smaller asteroids. Several tens of differentiated parent bodies are needed to account for the range of iron meteorite types (coming from the cores of such differentiated bodies) in our collections. However, a differentiated body of chondritic composition should be dominated by the olivine-rich mantle material (~60-80% in mass; Toplis et al., 2013). The apparent shortage of olivine-rich asteroids is a still poorly understood problem in solar system science. At least a dozen of alternative or complementary solutions have been proposed to this missing olivine problem. E.g., Consolmagno et al. (2015) suggested that early differentiated planetesimals may have had non-chondritic bulk compositions, hence not crystallising significant olivine in their mantles, while Elkins-Tanton et al. (2014) proposed that formation of significant olivine was prevented by high viscosity and rapid heat loss in planetesimals. We refer the reader to, e.g., Greenwood et al. (2015), and references therein, for a recent summary about the wide range of possible explanations for such underrepresentation of olivine-rich material that have been proposed. Here we stress that our results may support the “battered to bits” scenario (Burbine et al., 1996): mantle fragments from disrupted differentiated parent bodies could have been shattered to dimensions below a few hundreds of metres, below the limit of detectability of previous spectroscopic surveys. Noteworthy, our findings about the relative abundance of A-types among small asteroids may also strengthen the hypothesis of an exogenous origin for the olivine detected as isolated outcrops on the surface of Vesta (Turrini et al., 2016, and references therein).

In our survey we also found an unexpectedly high number of D-type asteroids, believed to be the most primitive in the solar system. Their very low albedo and featureless red spectra suggest a high abundance of organics and volatiles, which may hold clues regarding the planetary processes that preceded life on Earth (e.g., Hiroi et al., 2001). Indeed, isotopic ratio measurements seem to exclude the hypothesis of a significant contribution of comets to the water of the Earth’s oceans and atmosphere, and primitive, carbonaceous asteroids could have played the main role in this regard (Altwegg et al., 2015; Marty et al., 2016). However, D-type asteroids were considered to be very rare in the NEA population. Our results could conversely indicate that they are quite abundant among the small-sized bodies, probably as a consequence of the high fragility of these carbonaceous asteroids, which favors the fragmentation of larger bodies (e.g., Granvik et al., 2016). The robustness of this result is outlined by the fact that the very dark D-types should be more difficult to discover/observe at the smallest size ranges. Noteworthy, DeMeo et al. (2014) found a high abundance of D-types among the smallest (< 15 km) asteroids in the inner main belt, from which the NEA population is mostly replenished via well-known dynamical paths. Our results hence imply that the asteroidal contribution to the delivery of the prebiotic material to the primitive Earth could be even much more important than foreseen prior of our observations.

Being based on the observation of numerous different targets over a wide range of phase angles, our work also allows us to constrain the spectral phase reddening of asteroids belonging to different compositional/taxonomic groups. On average, we found that low-albedo asteroids (C-complex and D-types) show no/limited phase reddening (confirming and reinforcing the preliminary findings by Lantz et al. 2018), incidentally suggesting a brand new way to discriminate primitive objects within the X-complex, whenever measurements are available spanning a wide range of phase angles. We note however that a quite strong phase reddening (0.104±0.003 %/100 nm/deg) has been measured from observations of the nucleus of comet 67P/Churyumov-Gerasimenko by the OSIRIS instrument onboard the ROSETTA mission (Fornasier et al., 2015). The strongest phase reddening is seen for olivine-rich surfaces (A- and Q-types). This is in agreement with previous laboratory results showing that olivine-rich ordinary chondrites are the most affected by phase reddening, though the reason of such behaviour is still unclear (Sanchez et al., 2012). For all of the taxonomic groups, the dependence of the spectral slope by the phase angle is monotonous. We stress that the wavelength dependence of the brightness phase function has always been found monotonous for astronomical observations, with only a few exceptions: those of the martian surface by the Viking 1 lander evidenced a change from reddening to blueing with increasing phase angle, producing an “arch-shaped” phase dependence of the observed spectral slope (Guinness, 1981); moreover, a not-monotonous behaviour is displayed by some observations of the lunar soil (e.g., Shkuratov et al., 2011). Based on laboratory experiments and numerical simulations, Schröder et al. (2014) found a monotonous phase reddening for a microscopically rough regolith, and a non-monotonous arch-shaped behaviour for smooth surfaces. Grynko & Shkuratov (2008), using numerical modelling based on the geometric optics approximation, evidenced a monotonous phase dependence of the spectral slope in the case of single-particle scattering, and non-monotonous behaviour for multiple-component scattering. These authors also found that monotonous phase blueing can happen for particles larger than about 250 μm. Overall, our results hence suggest the presence of a microscopically rough regolith on the surfaces of the observed NEAs, and a similar surface texture for objects belonging to the same compositional group. We remind however that in our approach we averaged out other factors – e.g., peculiar space weathering and particle size – potentially affecting the spectral response of the surfaces of the individual asteroids in each taxon. More dedicated studies of the phase reddening (i.e., with a complete phase coverage for each object) are still needed.

As a concluding remark, we want to emphasize the very important role that the investigation of the “small” near-Earth asteroids can represent for solar system science. The results presented in this paper already provide some clues on different topics related to the formation and early evolution of planetesimals. We should however keep in mind that these results are based on a relatively limited quantity of available data. Future, larger-scale surveys for the systematic physical characterization of newly-discovered NEAs down to the metre-sized (possibly extending to near-infrared wavelengths, to minimize the risk of taxonomic misassignments) will be crucial to further test and extend our findings.

## Acknowledgements

This work is based on observations collected at the European Organisation for Astronomical Research in the Southern Hemisphere under ESO programme 095.C-0087. We acknowledge financial support from the NEOShield-2 project, funded by the European Union’s Horizon 2020 research and innovation programme (contract No. PROTEC-2-2014-640351). DP has received further funding from the Horizon 2020 programme also under the Marie Skłodowska-Curie grant agreement n. 664931. This work was also supported by the Programme National de Planétologie (PNP) of CNRS/INSU, co-funded by CNES.